\documentclass[runningheads]{llncs}
\usepackage[T1]{fontenc}
\usepackage{amssymb}
\usepackage[utf8]{inputenc}
\usepackage{graphicx}
\usepackage{multirow} 
\usepackage{subcaption}
\usepackage[inline]{enumitem} 
\usepackage{url}
\usepackage[utf8]{inputenc}
\usepackage[table,xcdraw]{xcolor}
\usepackage[noend]{algpseudocode}
\usepackage[linesnumbered, ruled]{algorithm2e}
\usepackage[hidelinks]{hyperref} 
\usepackage{array}
\usepackage[binary-units,per-mode=symbol,range-phrase=--]{siunitx}
\usepackage{amsmath}
\usepackage{amssymb}
\usepackage{mdwmath}
\usepackage{flexisym}
\usepackage[style=base]{caption}
\usepackage{makecell}
\usepackage{diagbox}
\usepackage{scalefnt}
\usepackage[inline]{enumitem}
\usepackage{pgfplots,filecontents}
\usepackage{comment}
\usepackage{footnote}
\usepackage{wrapfig}
\usepackage[switch]{lineno}
\pgfplotsset{compat=newest}
\usepgfplotslibrary{units}
\setcellgapes{1pt}
\usepackage{braket}
\usepackage{filecontents}
\usepackage{xcolor}
\usepackage{cite}

\usetikzlibrary{patterns}
\def\BibTeX{{\rm B\kern-.05em{\sc i\kern-.025em b}\kern-.08em
    T\kern-.1667em\lower.7ex\hbox{E}\kern-.125emX}}
\pgfplotsset{%
   every tick label/.append style = {font=\small},
   every axis label/.append style = {font=\small},
   every mark/.append style={scale=100},
   width=6.1cm,compat=1.3
}

\begin{document}

\mainmatter  

\title{Dynamic Resource Manager for Automating Deployments in the Computing Continuum}

\titlerunning{Resource Manager}

%

%
\author{Zahra Najafabadi Samani 
\and Matthias Gaßner \and Thomas~Fahringer  \and Juan~Aznar~Poveda \and Stefan~Pedratscher}
%
\institute{University of Innsbruck, Institute of Computer Science, Innsbruck, Austria.\\
\email{\{Zahra.Najafabadi-Samani, Matthias.Gassner, Thomas.Fahringer, Juan.Aznar-Poveda, Stefan.Pedratscher\}@uibk.ac.at}\\
}

\toctitle{Lecture Notes in Computer Science}
\tocauthor{Authors' Instructions}
\maketitle

\begin{abstract}
With the growth of real-time applications and IoT devices, computation is moving from cloud-based services to the low latency edge, creating a computing continuum. This continuum includes diverse cloud, edge, and endpoint devices, posing challenges for software design due to varied hardware options. To tackle this, a unified resource manager is needed to automate and facilitate the use of the computing continuum with different types of resources for flexible software deployments while maintaining consistent performance. Therefore, we propose a seamless resource manager framework for automated infrastructure deployment that leverages resources from different providers across heterogeneous and dynamic Edge-Cloud resources, ensuring certain Service Level Objectives (SLOs). Our proposed resource manager continuously monitors SLOs and reallocates resources promptly in case of violations to prevent disruptions and ensure steady performance. 
The experimental results across serverless and serverful platforms demonstrate that our resource manager effectively automates application deployment across various layers and platforms 
while detecting SLO violations with minimal overhead.


\textcolor{red}{Personal use of this material is permitted. Permission from Springer must be obtained for all other uses, in any current or future media, including reprinting/republishing this material for advertising or promotional purposes, creating new collective works, for resale or redistribution to servers or lists, or reuse of any copyrighted component of this work in other works.}
\end{abstract}

\section{Introduction} \label{sec:Introduction}
With the accelerating growth of IoT devices and real-time applications 
the landscape of computation is shifting from Cloud-centric services to a low-latency Edge, giving rise to the computing continuum~\cite{SurveyIEEE2021,Samani2023}. 
Managing resources within this computing continuum becomes particularly challenging due to the dynamic and heterogeneous nature of resources, resulting in inefficiencies in provisioning, allocation, monitoring, and potential violations of Service Level Objectives (SLOs), which current methodologies struggle to address effectively. Existing resource managers~\cite{Dustar2023, caballer2023infrastructure} often encounter limitations by relying on resources from a single Cloud vendor~\cite{aspir2022cross, Xiaodi21} or a specific resource type~\cite{Mehran2022, MRAPSamani2023}, restricting users from leveraging the advantages of alternative providers and nearby Edge devices with lower latencies. This paper introduces a dynamic resource manager framework for automated infrastructure deployment to negotiate and assign resources based on SLOs. The implementation leverages the reactive Java framework Eclipse \texttt{Vert.x} and follows a generic and reusable design. Our proposed resource manager involves continuous and dynamic monitoring of SLOs, and in the case of SLO violations, it swiftly notifies the client and reallocates resources to prevent disruptions and maintain consistent performance. Furthermore, the resource manager provides a user-friendly interface facilitating the interaction with the system. The contributions of the paper are as follows:

\begin{itemize}
    \item Manage dynamic Cloud and Edge resources including serverless function and containerized service platforms on top of Kubernetes.
    \item Providing resources across the computing continuum upon requested SLOs.
   
    \item Automatic deployments across different providers and platforms. 

    \item Continuously monitoring the performance, and availability of resources, networks, and running functions and services through various metrics. 

    \item Generating notifications upon predefined criteria to notify 
    automated systems of SLO violations and dynamic changes to adopt corrective actions. 
    \item Conducting detailed experiments and analyses throughout the real-world computing continuum, encompassing serverless and serverful platforms.
\end{itemize}

The paper has six sections. Section~\ref{sec:RelatedWork} surveys the related work. Section~\ref{sec:Arc} presents the architectural design of the resource manager. 
Section~\ref{sec:Evaluationsetup} provides the experimental setup. Section~\ref{sec:Evaluationresults} evaluates the proposed resource manager through different scenarios. Finally, Section~\ref{sec:Conclusion} concludes the paper.


\section{Related Work} \label{sec:RelatedWork}
This section reviews the resource managers within Edge, Cloud, and the computing continuum.
\subsection{Resource manager for Cloud}

Xiaodi et al.~\cite{Xiaodi21} developed a microservice-based resource manager 
to facilitate the integration of new features. Their method overcomes the bottleneck of the shared data store by using an in-memory data grid 
and improves availability using multiple schedulers. However, the system does not support SLO violations and various resource types, such as serverless computing.

Aspir~\cite{aspir2022cross} proposed a resource manager for the Cloud through a unification layer for various providers. Their method selects resources upon client constraints. 
However, it lacks support for serverless and containerized applications. 

\subsection{Resource manager for Edge}
Wang et al.~\cite{Wang2020Edgemanager} proposed a framework consisting of the node and server Manager to manage the resources in Edge computing. Their method processes requests by evaluating resource availability and request priority before initializing containers. 
However, it lacks monitoring and solutions for handling node failures.

Abuibaid et al.~\cite{Abuibaid2022} introduced a resource manager for Edge computing, 
to minimize the time required for failure recovery. Their method offers continuous monitoring and dynamic failover functionality, 
and efficient rescheduling. However, their technique does not include automated infrastructure deployment. 
\subsection{Resource manager for computing continuum}
Morichetta et al.~\cite{Dustar2023} proposed a framework to manage serverless functions in the computing continuum to fulfill SLO. They propose a platform considering a three-layer intent-based architecture that translates stakeholder requirements into actionable steps, to control the distributed infrastructure.

Caballer et al.~\cite{caballer2023infrastructure} introduced an infrastructure manager to 
automate the application deployment in the computing continuum. Their method utilizes the infrastructure as Code 
while leveraging Kubernetes. However, their method fails to identify and address SLO violations and application failures.

Jansen et al.~\cite{jansen2023continuum} introduced a framework to automate deploying infrastructure supporting a broad range of Clouds and Edge environments with diverse configurations. However, their method lacks monitoring mechanisms to address SLO violations and failures.

Samani et al.~\cite{Samani2021, MRAPSamani2023} introduced a scalable resource manager for the computing continuum to minimize resource wastage and service failures by monitoring SLOs and clustering resources based on network connectivity and computing features.  However, their method fails to support serverless functions.

Tusa et al.~\cite{tusa2023end}
introduced an orchestration framework to manage resources 
in the computing continuum. This method employs a graph-based model 
and select the resources through various algorithms and optimization goals. However, it does not provide monitoring of SLOs and is not suited for serverless functions.
\begin{description}[wide]
    \item[Limitation:] Most of these works are limited to Cloud or Edge. Additionally, only a few resource managers orchestrate across the entire computing continuum, and these typically support only specific types of resources. This restriction prevents users from fully utilizing the benefits of alternative platforms. 
\end{description}



\section{Architecture design}\label{sec:Arc}
\begin{figure*}[t]
    \centering
    \includegraphics[width=1\linewidth]{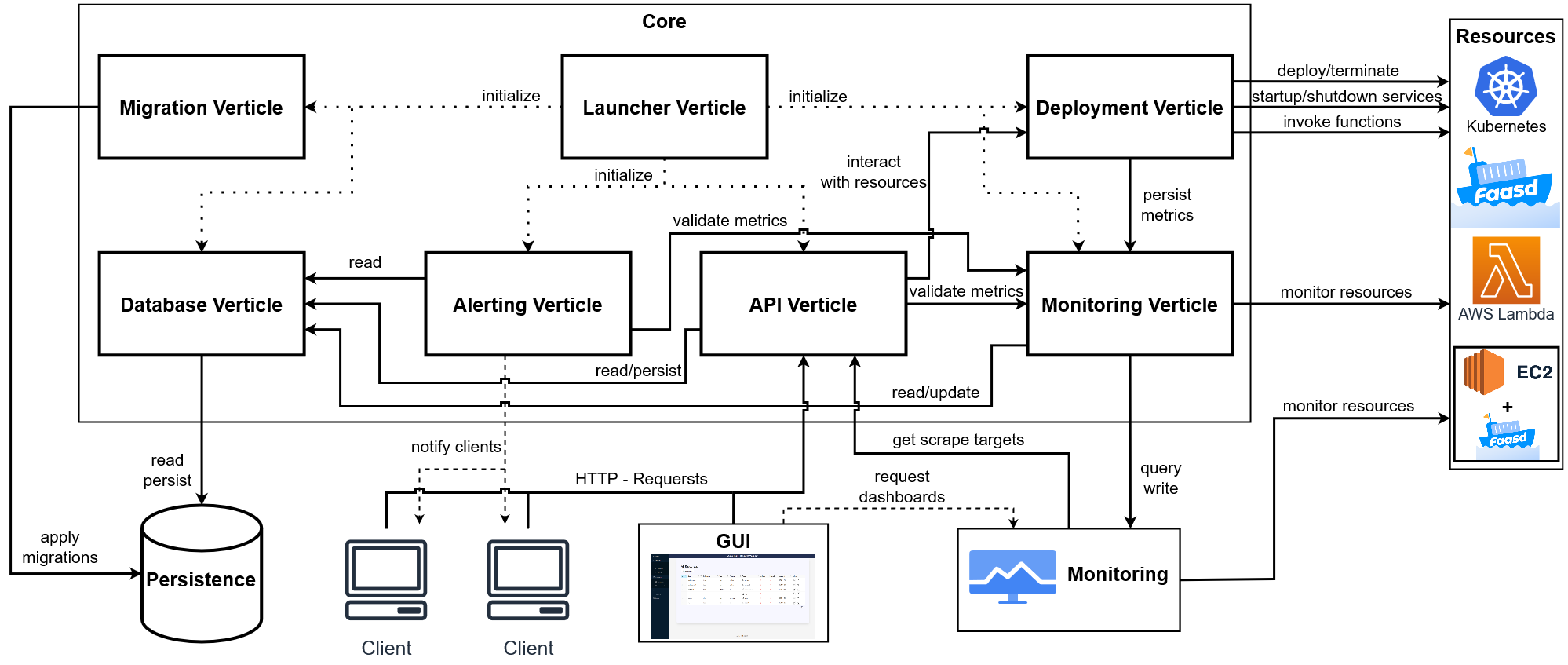}
    \caption{Architectural design}
    \label{fig:architecture}
\end{figure*}

This section provides an overview of the architectural design of the resource manager. The architecture mainly consists of four decoupled layers:  persistence,  core, GUI, and monitoring layers 
 (see Figure \ref{fig:architecture}). 

\subsection{\label{subsec:architecturepersistence}Persistence layer} 
The persistence layer ensures the 
persistence of user data, metadata of resources, services, and functions using a \textit{PostgreSQL} 
database.
It provides reliability and feature robustness. The database is only accessible by the core layer, which consists of database migrations as well as reading and writing the data. 
\subsection{Core layer}\label{subsec:architecturebackend}
The core layer serves as the backend of the resource manager and delivers essential functionalities via six key components called \textit{verticles}. It is developed using the \textit{Java} toolkit \textit{Eclipse Vert.x} \cite{vertx}. \textit{Vert.x} is characterized by its modular and extensible design, emphasizing a reactive programming model for efficient handling of concurrent operations. 
A verticle is a Vert.x specific object that implements an actor-like model consisting of multiple event loop threads that react to incoming events. All communications between verticles are realized through the event bus \cite{vertx-eventbus} and provide an API to send and receive simple messages or objects with publish/subscribe, point-to-point, and request-response messaging. Every verticle plays a distinct role in supporting a different aspect of the core functionality. The main features implemented by verticles are detailed as follows.

\subsubsection{Launcher Verticle}\label{susbsubsec:launcherverticle} serves as 
 the entry point of the resource manager. It is responsible for loading the global configuration, initializing verticles with the retrieved configuration, and subsequently activating them in a specific sequence. This sequential start-up is crucial to fulfill certain prerequisites for each verticle.

\subsubsection{Migration Verticle}\label{subsubsec:migrationverticle} is initiated by the Launcher Verticle and is responsible for all database migrations to the \textit{PostgreSQL} database. These migrations include creating new databases and updating or deleting existing ones. We used \textit{Flyway} \cite{flyway}, a widely used database migration tool, to manage the migrations.

\subsubsection{Database Verticle} \label{subsubsec:databaseverticle}
provides direct connectivity to the PostgreSQL database. 
This verticle is accessed by the \textit{API Verticle}, enabling database read and write operations. 
The \textit{Monitoring Verticle} relies on it to update information on resources and retrieve necessary data about monitoring targets. To enable alerting 
for SLO violation the \textit{Alerting Verticle} queries all resources registered for the validation of SLOs.
We utilize the \textit{Object-relational mapping (ORM)} tool \cite{hibernateorm} for the seamless mapping of \textit{Java} objects to database tables and vice versa. 

\subsubsection{Monitoring Verticle} \label{subsubsec:monitoringverticle}
is responsible for collecting metrics that the monitoring layer can not capture. 
It collects 
resource cost, 
latency and reachability of 
regions, 
and  \textit{Kubernetes} metrics exposed by the \textit{Metrics Server}. 
Monitoring Verticle gets metrics collected by the \textit{Deployment Verticle}. 
After collecting metrics, they are transformed into \textit{OpenTSDB} \cite{opentsdb} objects and pushed to the monitoring layer. 
Additionally, to validate the current state of resources, the \textit{Monitoring Verticle} queries the necessary metrics from the monitoring system, validates them and returns the result to the \textit{API} or \textit{Alerting Verticle}.


\subsubsection{Deployment Verticle} \label{subsubsec:deploymentverticle}
is responsible for the deployment and termination of resources triggered by the \textit{API Verticle}. Once a new deployment is created, it first validates the necessary resources and credentials. The deployment leverages \textit{Kubernetes} \cite{k8s}, 
providing effective management, deployment, and monitoring of resources across various environments. Additionally, the deployment process utilizes \textit{Terraform} \cite{terraform}, an Infrastructure as Code (IAC) tool, streamlining and automating the infrastructure setup. 
Deployment Verticle uses a pre-pull container that downloads and retains the Docker images on the system, using a tiny, paused image minimizing resource usage. 
Clients can manage their services through API calls, enabling them to start/ stop their deployments without re-pulling images by using the pre-pulling process. 
When clients no longer require the services, they terminate the deployment via a Rest API.

\subsubsection{API Verticle}
\label{subsubsec:monitoringverticle}
provides a Rest API for clients and a frontend to access the resource manager. The API is defined 
according to the \textit{OpenAPI Specification} \cite{openapi}, ensuring human and machine readability and a language-agnostic interface to reduce the complexity of application interactions. Additionally, the API Verticle enhances security by employing \textit{JSON Web Tokens (JWT)} \cite{jwt} for client authentication. This standard method securely transfers digitally signed data between parties. Tokens are issued only to clients with valid login credentials.

\subsubsection{Alerting Verticle} \label{subsubsec:alertingverticle} periodically checks for deployments with activated alerting and validates the resources of these deployments by communicating with the Monitoring Verticle. 
During the creation of a deployment with active alerting,
the client provides an HTTP endpoint capable of receiving a POST request. The resource manager utilizes this endpoint to dispatch alert notifications.


\subsection{Monitoring layer} 
\label{subsec:monitoring}
The monitoring layer is responsible for continuously observing the performance, health, and availability of resources, networks, and running functions and services. The monitoring system leverages \textit{Victoria Metrics} \cite{victoriametrics}, a monitoring solution 
known for its speed, cost-effectiveness, and scalability. It is 
compatible with various APIs from other prominent monitoring solutions such as \textit{Prometheus, Graphite, InfluxDB}, and \textit{OpenTSDB}.
Moreover, \textit{Victoria Metrics} adopts \textit{MetricsQL} as its primary query language, which serves as a superset of the widely used \textit{PromQL} query language, associated with \textit{Prometheus} \cite{prometheus}. This compatibility ensures 
the capability to scrape metrics from Prometheus metric exporters 
which are used in the implemented system to expose hardware- and kernel-related metrics from \textit{Amazon EC2} and \textit{OpenFaaS} resources. \textit{Victoria Metrics} offers advantages over \textit{Prometheus} due to its ability to serve as long-term storage for monitored data and its support for both pull and push models. 

\subsection{Graphical User Interface (GUI) layer} \label{subsec:architecturefrontend}
The GUI layer provides a user-friendly interface, facilitating interaction with the system. We created it utilizing the \textit{React} \cite{react} \textit{Next.js}, \cite{nextjs}, and the \textit{Ant Design} \cite{antd} framework. Additionally, we used \textit{tailwindcss} \cite{tailwind} as a CSS framework providing simple and efficient styling of the graphical user interface (GUI). Moreover, we embedded \textit{Grafana Dashboard} \cite{grafana} into the GUI to visualize the monitoring data. \textit{Grafana Dashboard} is an open-source monitoring platform that supports various data sources, including \textit{PostgreSQL} and \textit{Victoria Metrics}, providing a customizable and visually appealing interface for interactive dashboards.

\section{Evaluation setup
} \label{sec:Evaluationsetup}
This section provides the configuration details of our evaluation. We published the resource manager in the GitHub code repository~\cite{RMRepository}.
\subsection{Infrastructure}
We evaluated the resource manager on a computing continuum testbed consisting of serverless and serverful resources on top of K8s illustrated in Table~\ref{tab:evaluation_resources}. 
\begin{table*}
    \centering
    \resizebox{\textwidth}{!}{%
    \begin{tabular}{|c|c|c|c|c|c|c|c|c|c|c}
        \cline{2-10}
        \multicolumn{1}{c|}{}&\textbf{ID} & \textbf{Platform} & \textbf{Application} &\textbf{number} & \textbf{Resource}&\textbf{Location}&\textbf{CPU}&\textbf{Memory}&\textbf{Storage}\\
        \multicolumn{1}{c|}{}&&&\textbf{type}&&\textbf{type}&&\textbf{(core)}&\textbf{(GB)}&\textbf{(GB)}\\
        \hline
        \multirow{3}{.4cm}{\centering \textit{\rotatebox[origin=c]{90}{Edge}}}& 1 & OpenFaaS & FaaS & 1& Intel NUC& UIBK & 8   & 16   & 64\\
        \cline{2-10}
        & 2 & OpenFaaS & FaaS &  1& Raspberry Pi 3& UIBK & 4   & 1   & 16 \\
        \cline{2-10}
        & 3 & OpenFaaS & FaaS &  1 & Raspberry Pi 4& UIBK &4   & 4  & 32\\
        \hline
        \multirow{5}{.2cm}{\centering \textit{\rotatebox[origin=c]{90}{Fog}}} & 4-9 & K8s & Container & 6& K8s node& UIBK & 2-8    & 2-8 & 30   \\   
          &  & &  & & Proxmox machine &  &  &  &    \\ 
        \cline{2-10}
        & 10 & K8s & Container & 1& K8s control plane node& UIBK & 8   & 8 &160  \\
        &  & &  & 1&  Proxmox machine &  &    & &  \\
        \hline
        \multirow{3}{.2cm}{\centering \textit{\rotatebox[origin=c]{90}{Cloud}}}& 11 & AWS EC2 (+ OpenFaaS) & FaaS & 1& t2.large& US-east-1& 2   & 8  & 8 \\
        \cline{2-10}
        & 12 & AWS Lambda & FaaS & 1&Serverless& US-east-1 &1-6& 0.128&0.512 \\
        \cline{2-10}
        & 13 & AWS Lambda & FaaS & 1&Serverless& US-west-2&1-6& 0.128&0.512 \\
        \hline
    \end{tabular}
    }
    \caption{Computing continuum testbed.}
    \label{tab:evaluation_resources}
\end{table*}

\paragraph{\textbf{Edge computing:}} consists of two Raspberry Pi 3 and 4 and one Intel NUC (resources 1-3, see Table~\ref{tab:evaluation_resources}) as Edge devices located in the University of Innsbruck (UIBK). 
These devices run the \textit{OpenFaaS} runtime, \textit{FaasD}, and \textit{Node Exporter}, allowing them to manage function-as-a-service workloads and expose metrics.

\paragraph{\textbf{Fog computing:}} is located in UIBK and runs on a K8s cluster composed of seven nodes (resources 4-10, see Table~\ref{tab:evaluation_resources}), where different layers of the resource manager 
are deployed across resources 5 and 7 respectively. A Python application is also deployed within the K8s cluster (on resource 6) to measure the time and resource utilization.
    
\paragraph{\textbf{Cloud computing:}} consists of the following Amazon Web Services (AWS) EC2 and Lambda instances:
    \begin{itemize}
        \item {AWS EC2:} a virtual instance (resource 11), which uses the \texttt{t2.large} instance offering 2 vCPU cores and 8 RAM located in the \texttt{us-east-1}.
        \item {AWS Lambda:} Serverless services (resources 12 and 13) limited to 1024 MB of RAM and located in \texttt{us-east-1} and \texttt{us-west-2}.
    \end{itemize}

\subsection{Use-cases}
We used two Python functions and a service to evaluate the resource manager.

\begin{itemize}
    \item Function 1: sleeps for one second and sends a response back to the client.
    \item Function 2: immediately sends a response back to the client.
    \item Service 1: We used the nginx \cite{nginxCi} image as a containerized application. 
\end{itemize}

\subsection{Metrics}
The Resource Manager is assessed using the following metrics: 

\paragraph{\textbf{Deployment time}} is the time between when the resource manager completes validating a deployment request and when all resources associated with that request are ready for usage. The deployment includes starting and initializing \textit{AWS EC2} resources, uploading \textit{AWS Lambda} and \textit{OpenFaaS} functions as well as pulling container images onto \textit{K8s} resources.

\paragraph{\textbf{Termination time}} is the time between when the resource manager completes validating a termination request and when all resources are terminated.

    
\paragraph{\textbf{Response time}} is the time from when a client issues a request to create or cancel a deployment until the client receives a reply. 
    
\paragraph{\textbf{Resource utilization}} is the \textit{CPU} and memory usage during the creation and termination of deployments. 
    
\paragraph{\textbf{Reaction time}} represents the time between SLO violation and Resource Manager response time to this event. 
    
\paragraph{\textbf{Function invocation round trip time}} 
represents the total time required to invoke a function including the execution time of the function and the network transmission of the request and response.
    
\paragraph{\textbf{Startup and shutdown time}} is the time taken to start and shutdown a service. 

\subsection{Evaluation scenarios} 
We designed three scenarios to evaluate various metrics:

\paragraph{\textbf{Scenario 1}}
evaluates the time measurement and resource utilization for different resource compositions: \begin{enumerate*}
    \item Single OpenFaaS (resource 1, function 1), 
    \item Single Lambda (resources 12, 13, function 1),
    \item Single K8s (resource 5, service 1), 
    \item Single EC2 (resource 11, function 1), and 
    \item All resource types (resources 1, 2, 3, 5, 11,  12, 13, function 1, service 1). 
\end{enumerate*}

\paragraph{\textbf{Scenario 2}} evaluates the system reaction time in case of the SLO violation considering single, two, three, and four deployments with all resource types (resources 1, 2, 3, 5, 11, 12, 13, function 1, service 1).
We consider the latency set to less than ten seconds for SLO violation to ensure all resources fulfill it. 
To simulate the SLO violation we created a deployment with validation turned on. 

\paragraph{\textbf{Scenario 3}}
evaluates the round trip time of function 1 on resources 1, 2 (openfaas2), 11 (ec2), and 12 (lambda) and for up to 384 concurrent invocations. We compared our resource manager (RM) with the asynchronous client of the Python library \textit{httpx} \cite{pythonHttpx} (Python Client). 

\section{Evaluation results}
\label{sec:Evaluationresults}
\subsection{Scenario 1}
\begin{figure}
    \centering
    \includegraphics[width=\columnwidth]{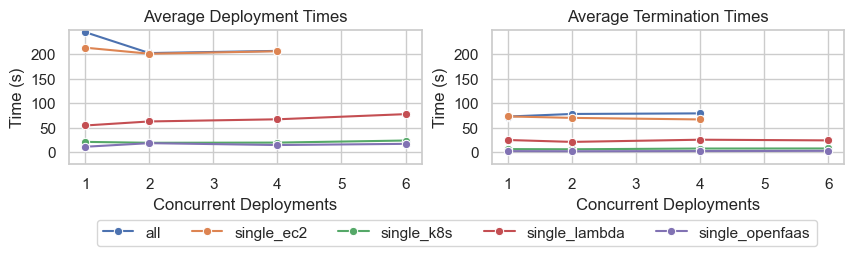}
    \caption{Average deployment and termination times for concurrent deployments}
    \label{fig:avgDeploymentTerminationTimes}
\end{figure}

\subsubsection{Average deployment and termination time} 

Figure \ref{fig:avgDeploymentTerminationTimes} shows that the deployment times are highly dependent on the type of resources. 
Single OpenFaaS deployments are the fastest followed by single K8s deployments. In both cases, the times slightly increase with more concurrent deployments because the target resource only needs to pull the corresponding Docker images and initialize a container. 
However, the deployment of lambda functions is significantly slower than the K8s and OpenFaaS deployments due to the higher network latency and a more complex setup mechanism of AWS. Again the necessary deployment time increases with more concurrent deployments, but the termination time stays the same.
The EC2 instance performs the worst, requiring approximately 200 s to start and about 65 s to terminate all deployments. This delay is due to the time-consuming processes involved in launching an EC2 instance, including setting up OpenFaaS and the Node Exporter, and higher network latency.

\begin{figure*}
    \centering
    \includegraphics[width=1\linewidth]{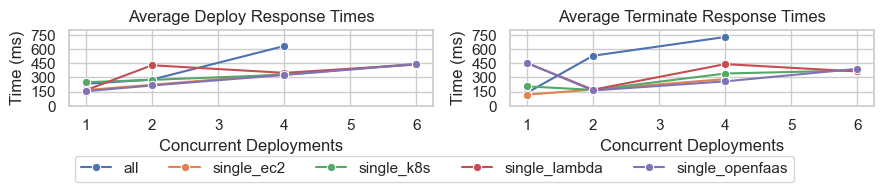}
    \caption{Average response times for concurrent deployments}
    \label{fig:avgResponseTimes}
\end{figure*}

\subsubsection{Response time}
Figure~\ref{fig:avgResponseTimes} shows the deployment and termination response time of the various types of resources for different concurrent deployments. Response times slightly increase in all cases with more concurrent deployments due to transaction isolation settings of the database. The database is configured to the serializable level to ensure that no two deployments lock the same resource simultaneously. 
In most cases, the deployment response time is similar to or slightly higher than the termination response time because deployments require some validation steps that are not necessary for the termination. The test case involving all resources shows the highest response time due to the increased validation needed when more resources are included in a deployment.

\begin{figure}
    \centering
    \includegraphics[width=\columnwidth]{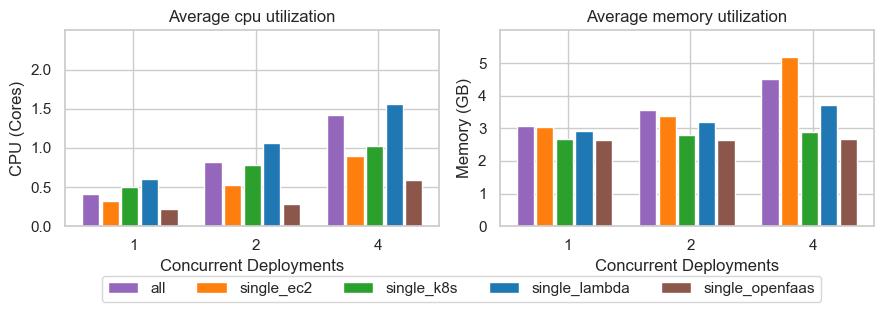}
    \caption{Resource utilization across various resource types}
    \label{fig:resourceUtilisation}
\end{figure}

\subsubsection{Resource utilization}
Figure \ref{fig:resourceUtilisation} shows resource managers' CPU and memory utilization across various resource types and concurrent deployments. 
OpenFaaS and K8s deployments have a lower impact on resource utilization because the Terraform providers used for these resources do not significantly affect memory and CPU usage. 
In contrast, deployments involving Lambda and EC2 resources utilize more memory and CPU due to AWS Terraform provider, which is more sophisticated and supports a broader range of services. 

\subsection{Scenario 2}
\subsubsection{Reaction time}
\begin{figure}
    \centering
    \includegraphics[width=0.4\columnwidth]{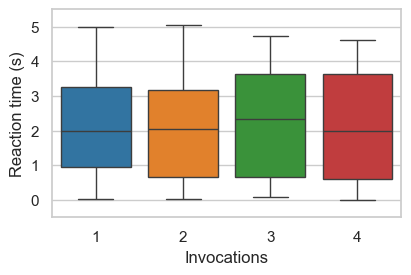}
    \caption{Reaction time for concurrent deployments}
    \label{fig:avgReactionTime}
\end{figure}

Figure~\ref{fig:avgReactionTime} shows the reaction time in response to the SLO violations. 
Reaction times consistently range between \num{0} and \num{5} s, as the evaluation interval for deployments is set to \num{5} s. The figure shows that even with an increase in the number of concurrent deployments, the reaction time remains unchanged, demonstrating that our system can swiftly detect SLO violations and respond effectively without being impacted by the increased deployment load.

\subsection{Scenario 3}

\subsubsection{Average function invocation round trip time}
\begin{figure}
    \centering
    \includegraphics[width=0.5\linewidth]{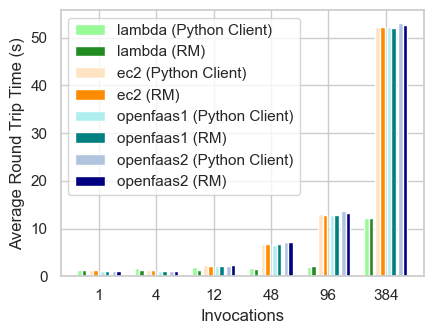}
    \caption{Average function invocation RTT} 
    \label{fig:avgFunctionInvocationTime}
\end{figure}
Figure~\ref{fig:avgFunctionInvocationTime} compares the round trip time (RTT) between our proposed resource manager and Python HTTP client
for various concurrent function invocations. Initially, the performance of the different resource platforms and invocation methods is quite similar for one and four concurrent invocations. However, as the invocations increase to twelve, there is a notable increase in the round trip time of EC2 and OpenFaaS resources compared to the AWS Lambda function. The reason is that the scalability of the EC2 and OpenFaaS resources is constrained by the number of CPU cores they possess, while the AWS Lambda function is designed to scale theoretically without such limitations.
In most cases, our resource manager has a lower round trip time, or roughly equivalent to that of the Python HTTP client. Therefore, we conclude that the resource manager does not add significant overhead to function invocations when compared to the \textit{Python} HTTP client.

\section{Conclusion} \label{sec:Conclusion}
This paper provided a resource manager framework on top of Kubernetes to automate infrastructure deployment in diverse computing environments. By leveraging the reactive capabilities of Eclipse Vert.x, the framework not only ensures a proactive approach to resource management but also introduces a robust system. The proposed manager effectively assigns resources to specified SLOs while continuously monitoring the availability of resources, networks, and running functions and services through various metric dashboards. We conducted extensive experiments and analyses throughout the real-world computing continuum, encompassing serverless and serverful platforms. The evaluation results reveal that our resource manager successfully automates application deployment across diverse layers and platforms within the computing continuum, and it detects SLO violations efficiently with minimal overhead.
\bibliography{main}
\bibliographystyle{unsrt}

\end{document}